\documentclass{svmult}
\usepackage{makeidx}  
\usepackage{graphicx}  
\usepackage{subfigure}   
\usepackage{epsfig}          
\usepackage{mathptmx,helvet,courier}  
\usepackage{multicol}        
\usepackage[bottom]{footmisc} 
\usepackage{cite}
\usepackage{amssymb} \usepackage{amsmath} \usepackage{amsfonts}

\def\rcgindex#1{\index{#1}}
\def\myidxeffect#1{{\bf\large #1}}

\begin{document}

 \title*{Experimental observation of moving intrinsic localized modes in germanium} 
  \titlerunning{Experimental observation of moving intrinsic localized modes in germanium}
  \author{Juan~F. R. ~Archilla \and
Sergio~M.M.~Coelho \and F.~Danie~Auret \and Cloud Nyamhere \and
Vladimir~I.~Dubinko \and Vladimir~Hizhnyakov
 } 
\institute{J.F.R.~Archilla
  \at Group of Nonlinear Physics, Universidad de Sevilla, ETSI Inform\'atica,
  Avda. Reina Mercedes s/n 41011, Seville, Spain, \email{archilla@us.es}
 \and S.M.M.~Coelho  and F.D.~Auret
 \at Department of Physics, University of
Pretoria, Lynnwood Road, Pretoria 0002, South Africa, \and
  C.~Nyamhere
  \at
  Physics Department, Midlands State University, P. Bag 9055, Gweru,
   Zimbabwe
 \and V.I.~Dubinko
  \at NSC Kharkov Institute of Physics and Technology, Akademicheskya Str.~1, Kharkov 61108,
  Ukraine
  \and V.~Hizhnyakov
  \at Institute of Physics, University of Tartu, Ravila 14c, 50411 Tartu,
  Estonia
  }
\authorrunning{JFR ~Archilla,
SMM~Coelho, FD~Auret, C. Nyamhere, VI~Dubinko, V~Hizhnyakov}

\tocauthor{J.F.R. ~Archilla, S.M.M.~Coelho, F.D.~Auret, C.
Nyamhere, V.I.~Dubinko and V. ~Hizhnyakov }

\maketitle \setcounter{minitocdepth}{1} \vspace{-3cm}\dominitoc
\newcommand{\De}{{\mathrm D}}
\newcommand{\hw}{\hbar\,\omega}
\newcommand{\nbr}{n_\b}
\newcommand{\Ea}{E_\mathrm{a}} 
\providecommand{\w}{\omega}
\newcommand{\Sb}{\text{Sb}}
\newcommand{\Ge}{\text{Ge}}
\newcommand{\cm}{\text{cm}}
\newcommand{\seg}{\text{s}}
\providecommand{\eV}{\text{eV}}
\newcommand{\nm}{\text{nm}}
\newcommand{\DB}{{\text{DB}}}
\newcommand{\ILM}{{\text{ILM}}}
\newcommand{\ph}{\text{ph}}
\renewcommand{\d}{\text{d}}
\newcommand{\kb}{k_B}
\newcommand{\RT}{\text{ph}}
\newcommand{\ann}{\text{ann}}
\newcommand{\app}{\text{app}}
\newcommand{\C}{$^\circ$C}
\newcommand{\Cmath}{^\circ\text{C}}
\vspace{0cm} \abstract{Deep level transient spectroscopy shows
that defects created by alpha irradiation of germanium are
annealed by low energy plasma ions up to a depth of several
thousand lattice units. The plasma ions have energies of 2-8\,eV
and therefore can deliver energies of the order of  a few eV to
the germanium atoms. The most abundant defect is identified as the
E-center, a complex of the dopant antimony and a vacancy with and
annealing energy of 1.3\,eV as determined by our measurements. The
inductively coupled plasma has a very low density and a very low
flux of ions. This implies that the ion impacts are almost
isolated both in time and at the surface of the semiconductor. We
conclude that energy of the order of an eV is able to travel a
large distance in germanium in a localized way and is delivered to
the defects effectively. The most likely candidates are
vibrational nonlinear wave packets known as intrinsic localized
modes, which exist for a limited range of energies. This property
is coherent with the fact that more energetic ions are less
efficient at producing the annealing effect. 
 } 
\section{Introduction}

In science like in many other aspects of human activity, there are
often fortunate coincidences that orientate research in unexpected
directions. In 2012 there was an international workshop in
Pretoria, South Africa, called NEMI 2012~\footnote{{NEMI 2012}:
1st {I}nternational {W}orkshop: Nonlinear effects in materials
  under irradiation, March 12-17, 2012, Pretoria, South Africa.
  P. Selyshev, chairman}. 
  Several
theoreticians and nonlinear physicists attended, among them there
were two of the authors. Several talks were intended for non
specialist in order that physics students could be able to
understand them. One of the subjects was nonlinear localized
excitations that travel along a periodic media without losing
energy and keeping their shape. They are called intrinsic
localized modes (ILMS) or discrete breathers (DBs). The first name
emphasizes the internal character of the phenomenon and reminds us
of the linear vibration modes or phonons. The latter name comes
from the observation of the internal vibration they experience
that can be compared with the breathing of a living being. They
were first obtained as an exact solution for the continuous
sine-gordon equation~\cite{germanium-sinegordon}. Simulations
using molecular dynamics are able to reproduce them in several
solids with energies of the order of a few tenths or a few units
of an eV.

Among the attendants was a PhD student, part of a research group
of the University of Pretoria working on defects in
semiconductors, particularly in germanium. They have obtained
unexpected results while treating Ge with low energy (2-8\,eV)
plasma ions. Those energies are known as subthreshold because the
threshold energy to produce displacements of atoms in germanium is
between 11.5 for the $\langle 111\rangle$ direction and 19.5\,eV
for the $\langle 100\rangle$
direction~\cite{germanium-holmstrom2010}. However, they had
observed that something was penetrating at least two $\mu$m inside
the germanium wafer and was able to anneal several defects, in
particular, the most abundant one, the E-center. The energy for
annealing an E-center is about 1.3\,eV, according to our
measurements and  theoretical
calculations~\cite{germanium-tahini2011}.  On the other hand the
maximum energy that an Ar ion of 4\,eV can transmit to a Ge atom
is 3.6\,eV, therefore the energies were precisely what was
expected for ILMs. A line of collaboration was started that joined
nonlinear theory, computer simulations, plasma physics and
semiconductor physics that eventually confirmed ILMs as the most
likely cause of the annealing~\cite{germanium-archilla-coelho2015}
and also suggested them as the explanation for other long-distance
effects such as the modification of defects by electron beam
deposition, where the energy transmitted was below
1.3\,eV~\cite{germanium-coelho2013}. In this workd we will try to
give an explanation of the different branches of the physics
involved and to analyze the reasoning that leads to the ILM
explanation and the consequences both for semiconductor physics
and nonlinear physics.
  \rcgindex{\myidxeffect{A}!Annealing of defects by ILMs}
  \rcgindex{\myidxeffect{D}!Defect annealing by ILMs}
  \rcgindex{\myidxeffect{A}!Annealing of defects in Ge}
    \rcgindex{\myidxeffect{D}!Defect annealing in Ge}
        \rcgindex{\myidxeffect{G}!Germanium (defect annealing by ILMs)}
\begin{figure}[b]
\sidecaption[b]
\includegraphics[width=6cm,clip]{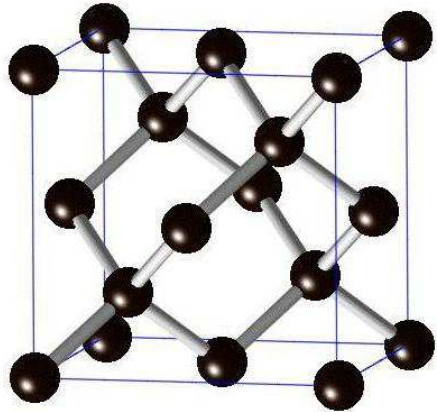}
\caption{Diamond structure of germanium. Each atom is bonded with
four nearest neighbours at the vertices of a tetrahedron. The
conventional cubic unit cell usually used is also shown. It
includes 8 atoms and can be seen as an fcc lattice with two atoms
at 0 and at 1/4 of the diagonal. The primitive cell has these two
atoms as a basis and the primitive vectors have their origin at 0
and end at the center of each adjacent face}
   \label{germanium-figure01} 
\end{figure}

\section{Germanium}
\begin{figure}[b]
 \begin{center}
\includegraphics[width=\textwidth,clip]{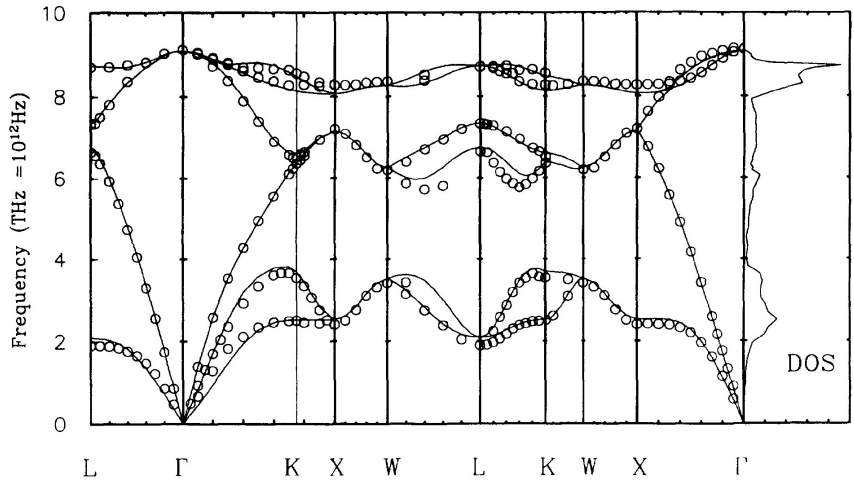}
\caption{Phonon dispersion and density of states for Ge.
Experimental values are shown as circles and theoretical
calculation are shown as solid lines. Modes about the center of
some optical bands with high frequency, large group velocity,
short wavelength and low dispersion may convert into ILMs when the
amplitude enters the nonlinear range. Reproduced with permission
from: Wei, S., Chou, M.Y.: Phonon dispersion of silicon and
germanium from first principles calculations. Phys. Rev. B
\textbf{50}, 2221 (1994). Copyright (1994) by American Physical
Society}
 \end{center}
\label{germanium-figure02}
\end{figure}

\begin{figure}[t]
 \sidecaption[t]
\includegraphics[width=6cm,clip]{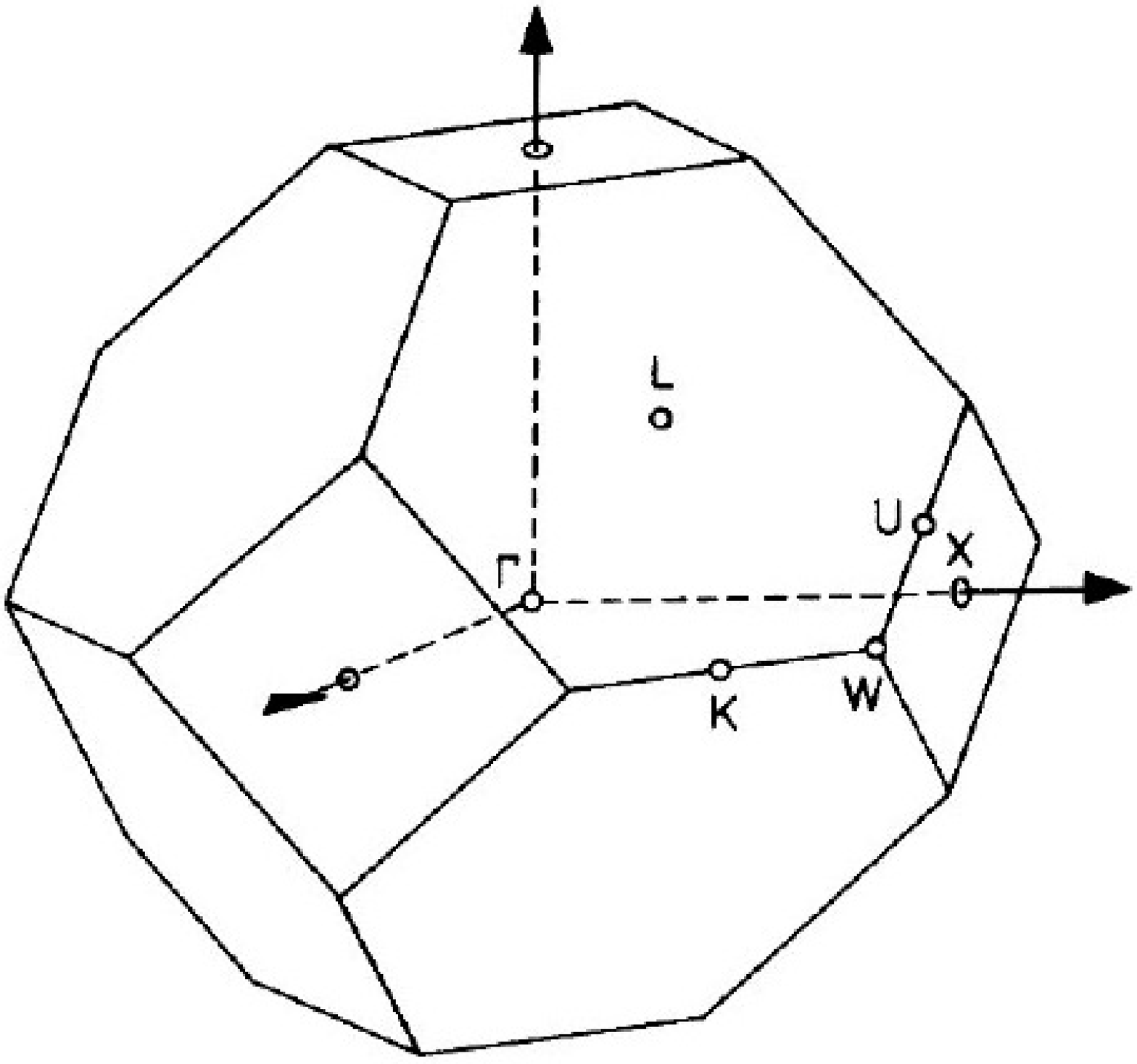}
\caption{Primitive Wigner-Seitz reciprocal cell for an fcc lattice
such as Ge, showing the directions in $k$-space and points that
appear in the spectrum shown in Fig.\ref{germanium-figure02}. The
point $\Gamma$ corresponds to wave number $k=0$, where the three
acoustical bands originate. The Wigner-Seitz cell is the region of
$k$-space that is closer to (0,0,0) than to any other point of the
lattice. Modes with wave vectors about the middle of $\Gamma$-L,
$\Gamma$-K and $\Gamma$-X may convert into ILMs when their
amplitude increases. Axes are the same as in
       \rcgindex{\myidxeffect{R}!Reciprocal cell of Ge}
              \rcgindex{\myidxeffect{G}!Germanium reciprocal cell}
Fig.~\ref{germanium-figure01}}
\label{germanium-figure03}
\end{figure}

       \rcgindex{\myidxeffect{D}!Diamond structure of Ge}
              \rcgindex{\myidxeffect{G}!Germanium diamond structure}

The diamond structure of germanium is well known, each atom has
covalent bonds with the four nearest neighbours at the vertices of
a tetrahedron as shown if Fig.~\ref{germanium-figure01}. Normally
a conventional cubic unit cell comprised of 8 atoms is used. The
diamond structure can be seen as an fcc lattice with two atoms at
points $(0,0,0)$ and at 1/4 of the
diagonal~\cite{germanium-ashcroft1976}. The lattice unit is
$a=5.66$\AA~for Ge, slightly larger than 5.43\AA~for Si and even
larger than 3.57\AA~for C diamond. The diamond structure is not
the best for moving ILMs because there is no chain of nearest
neighbours forming a straight line. This is a reason for which,
although stationary ILMs have been constructed with molecular
dynamics~\cite{germanium-voulgarakis2004}, the attempts to
construct moving ILMs have failed so far. Several lines of
research seem promising, one option is to construct ILMs in the
next neighbour directions such as $\langle 100\rangle$ where there
is a straight line of atoms. Other is to study  polarizations as
the ones observed for ballistic phonons in germanium or
silicon~\cite{germanium-northrop1980,germanium-karamitaheri2013}
which can travel distances of 160\,nm. It seems also possible that
ILMs can be nonlinear perturbations of linear optical modes with
high energy, high velocity, short wavelength and low dispersion,
such as at the middle of the  Brillouin  zone for optical
branches.
           \rcgindex{\myidxeffect{I}!ILMs in Ge}
           \rcgindex{\myidxeffect{G}!Germanium (ILMs in)
           }

The number of Ge atoms per unit volume can be obtained as
$n_\mathrm{Ge}=8/a^3=4.42\times10^{22}\,\mathrm{cm}^3$. Other
properties of interest are atomic number 32, atomic mass
$M=72.61$\,amu, density $\rho=5.323$\,g/cm$^3$, sound velocity
$c_s=5400$\,m/s,  Debye temperature $T_D=360$\,K, Einstein
temperature $T_E=288$\,K, covalent radius 1.22\,\AA, atomic radius
1.52\,\AA, melting point 1210.55\,K, 1st ionization energy
7.899\,eV and  specific heat 0.32\,J/gK at 300\,K.
              \rcgindex{\myidxeffect{G}!Germanium properties}
              \rcgindex{\myidxeffect{P}!Properties of germanium}
\section{Phonons in \textrm{Ge}} \label{germanium-sec:phonons} The
objective of this subsection is to review the well known concepts
of lattice dynamics and to see how they apply to Ge and to justify
subsequent calculations. Phonons are the usual means for energy
transport in a crystal and the responsible party for thermal
annealing of defects. With this review we want to demonstrate that
they cannot be th responsible for the annealing of the E-center
defect during Ar plasma bombardment. We will frequently use
general concepts of lattice dynamics and the reader can consult
any textbook, for example
Refs.~\cite{germanium-ashcroft1976,germanium-dove}.


In classical mechanics for a crystal with $n_\mathrm{at}$ per unit
volume, there are $3n_\mathrm{at}$ degrees of freedom. In the
harmonic approximation the substitution of
$u_{k,\w}=\vec{A}\exp(\I \vec{k}\cdot\vec{r}-\w t)$ in the
equation of movement leads to $3n_\mathrm{at}$ different linear
modes of frequency $\w$, wave number $\vec{k}$, phase velocity
$c=\w/k$ and polarization $\vec{A}$. They are organized in
branches $\w=\w(k)$, three of them are acoustic, that is $\w$
vanishes linearly with $k$ in the long wavelength limit. If the
crystal has a basis of $p$ atoms or ions in each primitive cell,
there are also $3(p-1)$ optical branches, that are bounded from
below. In Ge with two Ge atoms in the unit cell, there are three
optical branches. Each branch has $n_\mathrm{Ge}/2$ modes.
                   \rcgindex{\myidxeffect{P}!Phonons in germanium}
                     \rcgindex{\myidxeffect{G}!Germanium (phonons in)}
In the classical description, each mode can have any energy $E$
with a probability at temperature $T$ given by Maxwell-Boltzmann
equation $P(E)=\exp(-E/k_BT)/k_BT$, which leads to an average
energy $k_B T$ that is identical for each mode. Therefore, it is
trivial to obtain the energy per unit volume $u=3k_BT
n_\mathrm{Ge}$ and the specific heat at constant volume
$c_V=\partial u/\partial T=3k_Bn_\mathrm{at}$, a result known as
the Dulong-Petit law. This result is approximate at room
temperatures and above but fails spectacularly at lower
temperatures, which led to the quantum description of the harmonic
crystal. The classical description of the linear modes of the
crystal remains valid but the statistics are quite different.
                   \rcgindex{\myidxeffect{C}!Classical phonons}
                    \rcgindex{\myidxeffect{Q}!Quantum phonons}

In quantum mechanics a linear oscillator with frequency $\w$ can
only have energies given by $E_n=\frac{1}{2}\hbar\w+n\hbar\w$,
where $n$ is the excitation or occupation number. As the ground
state energy $\frac{1}{2}\hbar\w$ cannot be lost we will usually
suppress it and use
 \begin{equation}
 E_n=n E,\quad \mathrm{with}\quad E=\hbar\w\,,
 \label{germanium-eq:ehw}
 \end{equation}
where $E=\hbar\w$ is the quantum of energy, also called the energy
level.

At a given temperature $T$, the average values  $\langle n\rangle$
and $\langle E_n\rangle$ can be obtained with Bose-Einstein
statistics. They are
\begin{equation}
 \langle n\rangle=\frac{1}{\E^{E/k_bT}-1}\,\quad,\quad
  \langle E_n\rangle= \langle n\rangle E=\frac{E}{\E^{E/k_bT}-1}\,,
 \label{germanium-eq:bosen}
 \end{equation}
where $k_B= 8.617\times 10^{-5}$\,eV/K is the Boltzmann constant.

In a solid with $3n_\mathrm{at}$ degrees of freedom and therefore
the same number of linear modes, each one is equivalent to a
linear oscillator with a given frequency $\w$. It is usual to
describe them as phonons or quasi-particles and to use the
expression  $n$ phonons of a particular type with energy
$E=\hbar\w$ instead of a linear mode or state with frequency $\w$
and excitation number $n$. We will also follow this convention
although in some instances it may be more convenient to revert to
the original terminology.
       \rcgindex{\myidxeffect{N}!Number of phonons}
           \rcgindex{\myidxeffect{P}!Phonon statistics}
\begin{figure}[t]
 \begin{center}
\includegraphics[width=8cm,clip]{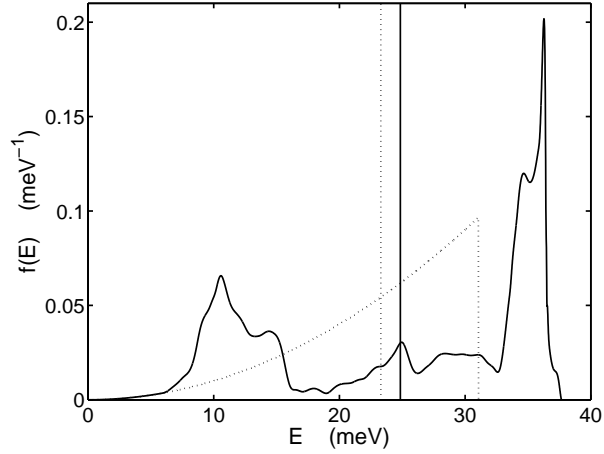}
\caption{Comparison with the number function or normalized density
of states for germanium: obtained from
Fig.~\ref{germanium-figure02} and Ref.~\cite{germanium-wei1994}
(---); Einstein model with $T_E=288$ (vertical ---) ;  Debye model
with $T_D=360$ ($\cdots$). The mean mode energy is for the first
two $E_\textrm{ph}=k_B T_E\simeq 25$\,meV. The Debye model by
definition has a maximum energy $K_BT_D=31$\,meV with average
energy 23.3\,meV equivalent to 270\,K (vertical $\cdots$) smaller
than the Einstein's one. The acoustic and optical bands can be
seen although they overlap}
 \end{center}
\label{germanium-figure04}
\end{figure}

As the number of frequencies is very large and they are very
close, $\w$ and $E=\hbar \w$ are considered as quasi-continuous
variables. Most energy levels are degenerate, i.e., there is more
than one mode with that energy, and in the  quasi-continuous
description there are very many in an interval $[E,E+\D E]$.

A variable density of states (DOS) $g(E)$ is introduced, also some
times called the density of levels. It is defined such as $g(E)\D
E$ is the number of linear modes or quantum phonon states per unit
volume with energies between $E$ and $E+\D E$. For a discrete
system the phonon spectrum is always bounded from above, that is,
there is a maximum frequency and energy $\w_M$ and
$E_M=\hbar\w_M$, therefore
 \begin{equation}
  \int_0^{E_M} g(E)\D E=3n_\textrm{at}\,.
 \label{germanium-eq:normge}
  \end{equation}
A rough estimate of the maximum value of energy level for the
acoustic modes can be obtained using the fact that the minimum
value of the wavelength is twice the lattice unit of the primitive
cell $d_{a}$, then $E_\mathrm{M,ac}\simeq \hbar \w_{M,ac}=\hbar c
2\pi/2d_{a}$, with $c$ the speed for the mode. For Ge,
$d_a=a/\sqrt 2=4.00$\,\AA\, and using $c_s=5400$\,m/s, we obtain
$E_\mathrm{M,ac}=28$\,meV and $f_\mathrm{M,ac}=6.7$\,THz similar
at the observed values in Figs.~\ref{germanium-figure02} and
\ref{germanium-figure04}. However, such a simple estimate for the
optical modes is not possible because the phase velocity tends to
infinity when $k\rightarrow 0$.
                   \rcgindex{\myidxeffect{D}!Density of states for phonons}
                   \rcgindex{\myidxeffect{P}!Phonon density of states}

Generally speaking there is no minimum frequency or energy as
explained above, however, when considering only a part of the
system, it can be described as subjected to an external potential
representing the interaction with the rest of the crystal. In this
case the phonon spectrum becomes optical, i.e., bounded from
below.

The energy of the solid per unit volume is given by
\begin{equation}
  u_E=\int_0^{E_T} \langle n(E)\rangle E \,g(E)\D E\,=\int_0^{E_T} \frac{E }{\E^{E/k_bT}-1}g(E)\D
  E\,.
\label{germanium-eq:evolume}
 \end{equation}

  We will also use the number density or
normalized density of states $f(E)=g(E)/3 n_\Ge$, with the
property that as $f(E)\D E$ is the fraction of modes with energies
between $E$ and $E+\D E$ and therefore the normalization condition
and average phonon energy $E_\mathrm{ph}$ are given by
\begin{equation}
\int_0^{E_T} f(E)\D E=1 \quad,\quad E_\mathrm{ph}=\int_0^{E_T}E
f(E)\D
  E\,.
 \label{germanium-eq:ephonon}
  \end{equation}

\begin{figure}[t]
\begin{center}
\includegraphics[width=5.7cm,clip]{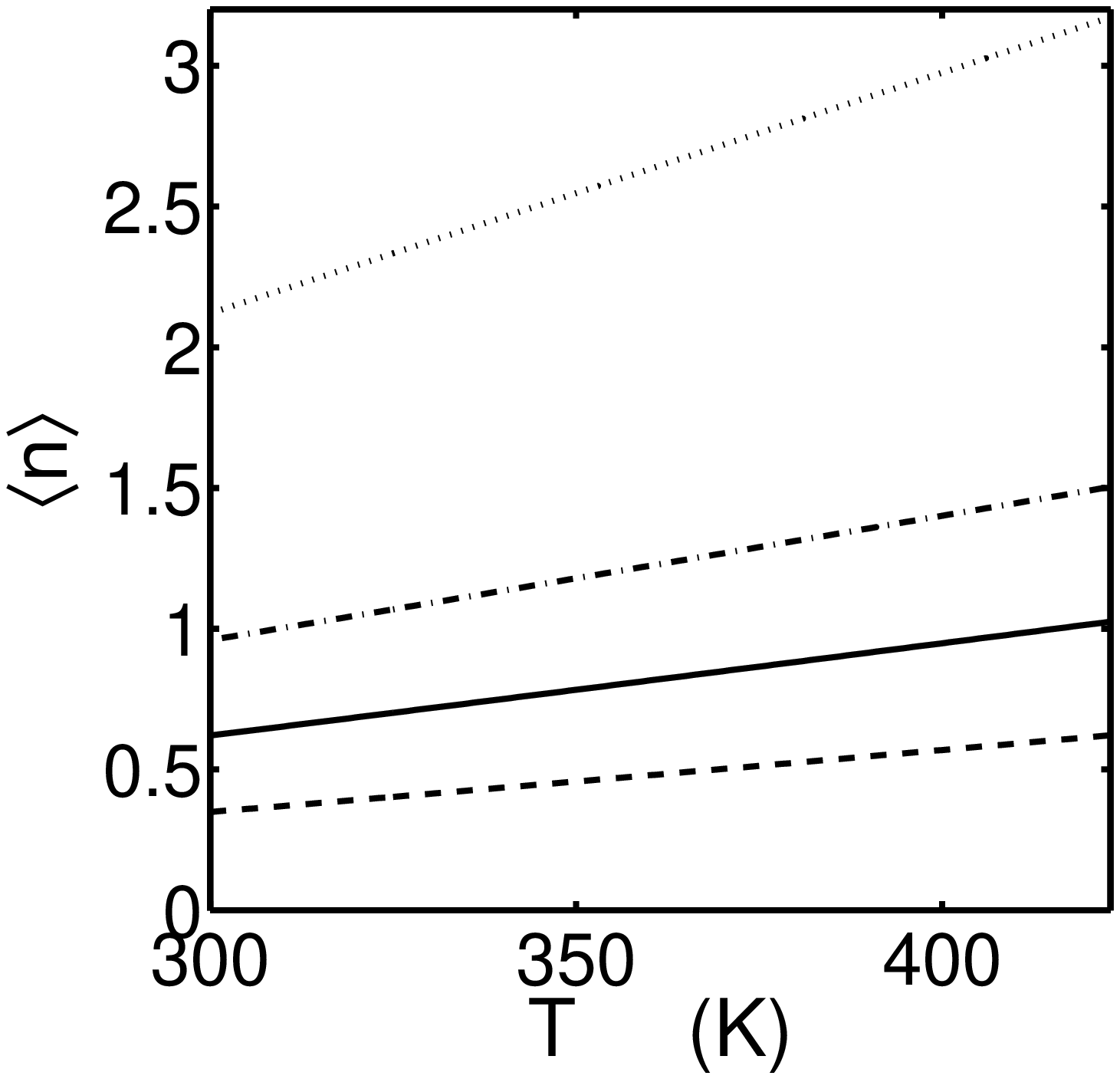}
  \hfill
\includegraphics[width=5.7cm,clip]{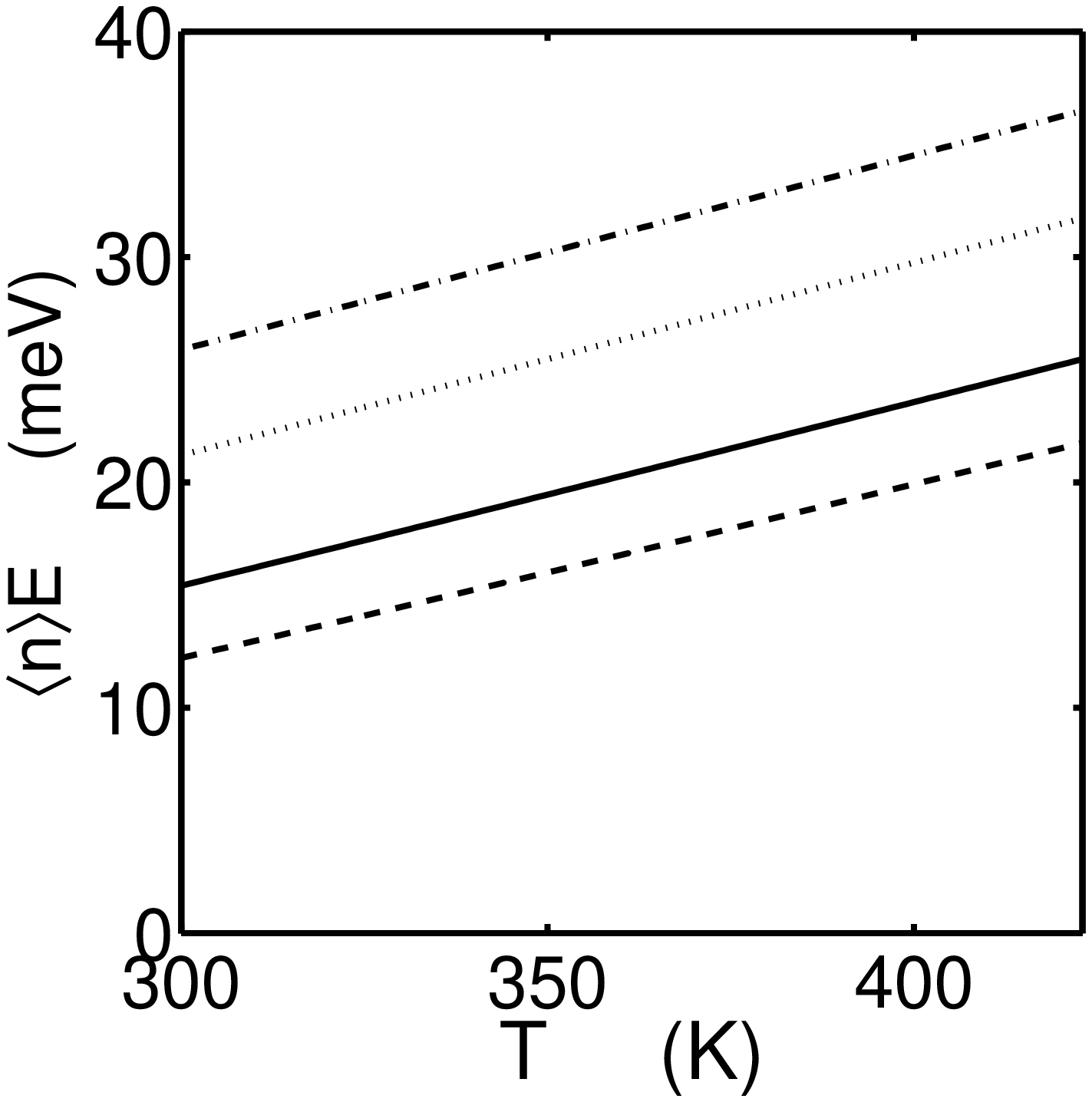}
\caption{(Left): Average number of phonons with respect to
temperature between room temperature $T_R=300$\,K and E-center
annealing temperature $T_A=423$\,K. From bottom to top: optical
phonons with $E_\mathrm{op}=35$\,meV; Einstein phonons with
$E_E=24.9$\,meV; average number of phonons with Ge DOS; acoustic
phonons with $E_\mathrm{ac}$=10. (Right): Average energy for
different phonons, from bottom to top:  optical phonons with
$E_\mathrm{op}=$35\,meV; Einstein phonons with $E_E=24.9$\,meV
($T_E=288$\,K) indistinguishable from the one obtained with Ge
DOS; acoustic phonons with $E_\mathrm{ac}=10$\,meV and average
classical energy $k_BT$. It can be seen that the acoustic modes
have more phonons and more energy than the optical ones and at
room temperature and above a quantum description is necessary}
 \end{center}
\label{germanium-figure05}
\end{figure}

There are two approximations frequently used for the density of
states: the Debye and the Einstein models. In the Debye model, all
phonon modes are substituted by three acoustic branches with
dispersion relation $\omega=c k$, with the same  $c$,  which is an
average velocity. These acoustic branches lead to a density of
modes or states per unit volume $g_D(E)=3/(2\pi^2\hbar^3 )c^3
E^2$~\cite{germanium-ashcroft1976}. Then,
$f_D(E)=g(E)/n_\mathrm{at}=\alpha_D E^2$, with the constant
$\alpha_D$ depending on the particular solid through $c$ and
$n_\mathrm{at}$. The energy has a cutoff value $E_D$ such that the
condition of normalization $\int_0^{E_D}f_D(E)\D E=1$ is
fulfilled.  Therefore, $\alpha_D E_D^3/3=1$. The values $E_D$ and
$T_D=E_D/k_B$ are known as the Debye energy and temperature,
respectively. Therefore there is only one unknown, either  $c$ or
$T_D$, either of which cannot be measured as they do not
correspond to real magnitudes. What is done is to choose $T_D$
such that the specific heat $c_v(T)$  fits the measurements. For
Ge, a value of $T_D=360\,K$ or $E_D= k_B T_D=31.1$\,meV is usually
given, which corresponds to $c=3420$\,m/s. This velocity is not a
real quantity but coherently it is approximately the mean of the
velocities of the two transversal modes, $\simeq 2500$\,m/s, and
the longitudinal one, 5400\,m/s~\cite{germanium-lacroix2005}. The
Debye dispersion relation works, of course, better for the
acoustic branches and small wave vectors.

\begin{figure}[b]
\begin{center}
\includegraphics[width=8cm,clip]{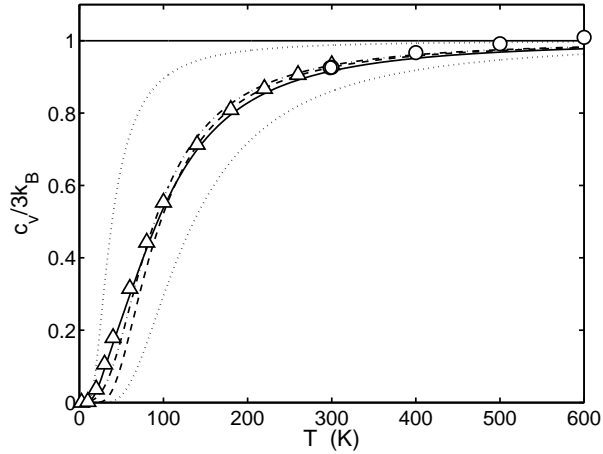}
\caption{Comparison of the experimental specific heat per degree
of freedom: (--) using germanium density of
states~\cite{germanium-wei1994}; (-\,-) the Einstein model with
$T_E=288$; ($-\cdot-$) the Debye model with $T_D=360$; ($\circ$)
and ($\Delta$) experimental values from
Refs.~\cite{germanium-berger1996,germanium-lide2010},
respectively. The horizontal line corresponds to the classical
Dulong-Petit law. The Einstein and the Debye model are slightly
better at intermediate temperatures, because they have been fitted
for that. At high temperatures the experimental $c_v$ becomes
larger because actual frequencies also increase with temperature.
The two separated dotted curves correspond to two Einstein models
with energies $E_\mathrm{ac}=10$ and $E_\mathrm{op}=35$\,meV, the
upper and lower curves, respectively. These values are
representative of the acoustic and optical branches}
 \end{center}
\label{germanium-figure06}
\end{figure}


The Einstein model supposes that there are $3n_\Ge$ modes with the
same  frequency $\omega_E$, being $E_E=\hbar\omega_E$ and
$T_E=E_E/k_B$, the Einstein energy and temperature, respectively.
The value of $T_E$ is chosen so as to fit the specific heat of the
solid, being $E_E$ an average energy of the phonons in the
crystal. For germanium its value is $T_E=288$\,K and will be used
in this work. In this model the mean energy per unit volume at
temperature $T$ in germanium is  simply
\begin{equation}
u_E=\frac{3n_\Ge}{\E^{E_E/k_B T}-1}
 \label{germanium-eq:uephonon}
\end{equation}


The actual phonon dispersion relation and the density of states
have been obtained and checked with experimental ones in
Ref.~\cite{germanium-wei1994}. Both magnitudes are shown in
Fig.~\ref{germanium-figure02}. The normalized density of states
$f(E)$ can be obtained from it but as the resolution is poor for
low energies we have substituted that part by the Debye one. The
Ge density of states is shown in Fig.~\ref{germanium-figure04}
together with the corresponding one for the Debye and Einstein
model for comparison. For $g(E)$ two concentrations of states
appear near the top and near the bottom of the spectrum, with a
drastic simplification we can describe them as an optical band
around $E_\mathrm{op}=35$\,meV and an acoustic one around
$E_\mathrm{ac}=10$\,meV. The mean phonon energy
$\int_0^{E_T}f(E)E\D E$ is approximately equal to the Einstein
energy.

Figure~\ref{germanium-figure05} represents the number of phonons
and the average energies as a function of temperature for acoustic
phonons, optical phonons, Einstein phonons, and average values
obtained with the density of states $g(E)$. It can be seen that
the classical statistics is not valid at the temperatures of
interest in this work and that there are significant differences
between optical and acoustic phonons. The energy in the acoustic
modes is larger than in the optical ones in spite of having less
energy but with more phonons. It can also be seen that the average
number of phonons $\langle n\rangle $ is smaller or closer to one
which indicates that the classical description is not good at room
temperatures and above.

Figure~\ref{germanium-figure06} represents the specific heat at
constant volume obtained from these models. There is no
significant  difference at the temperatures of interest in this
work between room temperature $T_R=300$\,K and the annealing
temperature of the E-center $T_A=423$\,K. This justifies the use
of the Einstein density of states  as a good approximation for
calculations. The specific heats for two Einstein models with
$E_\mathrm{ac}$ and $E_\mathrm{op}$ are also represented for
comparison.

\section{Defects and their detection with DLTS}

       \rcgindex{\myidxeffect{D}!Defects in germanium}
              \rcgindex{\myidxeffect{G}!Germanium defect detection}
               \rcgindex{\myidxeffect{D}!Defect detection with DTLS}
              \rcgindex{\myidxeffect{D}!Deep level transient spectroscopy (DLTS)}
Point defects in the structure or the type of atoms of the
semiconductor can appear with some probability due to the
temperature but they can also be created by radiation. In the
experiments described in this work most of the defects are created
by 5\,MeV  alpha
radiation~\cite{germanium-kolkovsky2007,germanium-roro2009}
produced in the decay of the
americium isotope $^{241}$Am.  A Ge sample 
 with dimensions $3\times 5\times 0.6$\,mm is
 brought into contact with americium
foil for 30 minutes.
\begin{figure}[t]
 \begin{center}
\includegraphics[width=9cm]{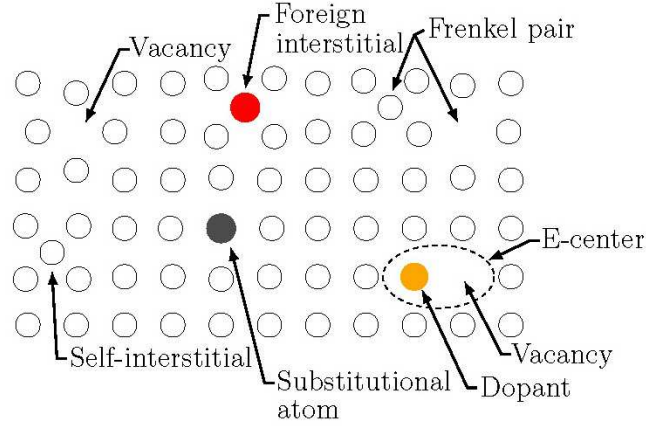}
\caption{Examples of some point defects in a crystal. The E-center
studied in this work is a complex of a vacancy and a
substitutional dopant Sb \label{germanium-figure07}} 
 \end{center}
\end{figure}

Defects can be of many types, some simple examples are shown in
Fig.~\ref{germanium-figure07}, such as a vacancy, a substitutional
atom, a self-interstitial, a foreign interstitial,  a Frenkel pair
that is, a combination of a vacancy and a self-interstitial and an
E-center, which is a combination of a dopant substitutional atom
and a vacancy. The germanium sample used in this work is doped
with antimony (Sb), with a dopant concentration
$n_\mathrm{Sb}=1.3\times 10^{15}\mathrm{cm}^{-3}$. Dopant atoms as
Sb atoms occupy substitutional positions but are not considered
defects as they are an essential part of the semiconductor
electrical properties. The main defect appearing after $\alpha$
irradiation is the E-center already described. There are many
others types such as vacancy complexes like the di-vacancy (V-V),
the tri-vacancy (V$_3$), the tetra-vacancy (V$_4$) and
combinations of interstitials as di or tri-interstitials (I$_2$,
I$_3$). Also, hydrogen (H), due to its small size is able to
penetrate almost everywhere and can combine with other defects
forming complexes such as VH$_n$, where n is an integer with
values from $1-4$. A variant of the E-center is the A-center, a
complex of an oxygen interstitial and a vacancy.
      \rcgindex{\myidxeffect{E}!E-center defect}
  \rcgindex{\myidxeffect{D}!Defect (E-center)}

Defects can experience many processes like diffusion, interaction
between them, modification, annealing and others. Generally
speaking all these processes are enhanced by temperature and the
rate at which the process takes place depends in an Arrhenius form
on a quantity known as the enthalpy for the process or sometimes
referred to as the activation energy or barrier energy for the
process. That is
\begin{equation}
\kappa\propto \E^{ -\Delta H/k_B T}\,.
\end{equation}

 The semiconductor Ge has a band gap of $E_g=0.67$\,eV. Some
defects introduce electrical levels inside the band of a
semiconductor, as for example in Sb-doped Ge, Sb introduce levels
very close to the conduction band.  When they are within the band
gap and more than 0.1\,eV  from the conduction or the valence
bands they are considered {\em deep}. Usually they are called
electron {\em traps} when they introduce an electron level and
hole traps when they introduce a hole level, respectively.  We
will write only about electron traps for simplicity, because the
treatment of holes is very similar, and because the main defect we
are interested in, the E-center, is an electron trap. The E-center
is located at $E_T=0.38$\,eV below the conduction band. The same
defect has also been reported as $E_T=0.37$.

When an electron is in a trap level it has a mean time of
permanence $\tau_n$ and its inverse $e_n=1/\tau_n$ is the
probability of emission per unit time. This magnitude and its
dependence on temperature are key to defect detection as it is the
actual magnitude measured in
DLTS~\cite{germanium-Lang1974,germanium-schroder2006}. This
dependence can be easily deduced.
           \rcgindex{\myidxeffect{E}!Electron trap}
              \rcgindex{\myidxeffect{T}!Trap of electrons}

Suppose that there are $N_T$ traps per unit volume, the
probability for an electron occupying the trap level of energy
$E_t$ (not $E_T$ which is $E_T=E_c-E_t$) is given by the
Fermi-Dirac distribution
\begin{equation}
f_t=\frac{1}{\E^{(E_t-E_F)/K_BT}+1}\,, \label{germanium-eq:fermi}
\end{equation}
where $E_F$, the Fermi energy is located near the middle of the
phonon band.

The probability that a moving electron is captured by a trap is
given by $c_n=\sigma_\mathrm{app}v_\mathrm{th}N_T(1-f_t)n$, where
$\sigma_\mathrm{app}$ is the capture cross section of an electron
for the trap, $v_{t}h$ is the thermal velocity of the electrons,
$N_T$ the trap concentration, $(1-f_t)$ the probability of the
trap being empty and $n$ the number of electrons per unit volume.
The latter quantity can be obtained as
$n=N_c\exp(-(E_c-E_F)/k_BT)$, where $E_c$ is the bottom energy of
the conduction band, $m_e^*$ being the effective mass of an
electron and $N_c=2\left(2\pi m_e^*k_B T/h^2\right)^{3/2}$ is the
effective density of states in the conduction
band~\cite{germanium-ashcroft1976}. The thermal velocity can also
be obtained as $v_\mathrm{th}=(2E_{th}/m_e^*)^{1/2}$, with
$E_\mathrm{th}=3/2k_BT$.
          \rcgindex{\myidxeffect{C}!Capture rate of electrons}
              \rcgindex{\myidxeffect{E}!Electron capture rate}

The trap emission rate $r_n$ is given by $r_n=N_T f_t e_n$, that
is, the concentration of traps multiplied by the probability of
being occupied and the probability of emission per unit time for a
trap. At thermal equilibrium $c_n=r_n$ and $e_n$ can be isolated
as
 \begin{equation}
   e_n =\sigma_\text{app}N_c v_\text{th}\exp(-E_T/\kb T)\,,
 \label{germanium-eq:emissionrate}
 \end{equation}
with $E_T=E_c-E_t$, that is, the distance of the trap level to the
conduction band.
         \rcgindex{\myidxeffect{E}!Emission rate of electrons}
              \rcgindex{\myidxeffect{E}!Electron emission rate}

It is easy to check that the pre-exponential factor is
proportional to $T^2$ as the effective mass is approximately
constant at the bottom of the conduction band where most of the
occupied states are.

Some authors discuss the interpretation of this expression of the
emission rate ~\cite{germanium-dimitrijev2009} as a function of
the capture parameters, however $\sigma_\mathrm{app}$ and  $E_T$
are considered the defect signature and used worldwide.
Independently of the meaning $\sigma_\mathrm{app}$ has the right
dependence on the temperature and should simply considered simply
as a parameter of the defect.
           \rcgindex{\myidxeffect{C}!Capture cross section of electrons}
              \rcgindex{\myidxeffect{E}!Electron capture cross section}
              \rcgindex{\myidxeffect{C}!Cross section for electron capture}

                 \rcgindex{\myidxeffect{D}!DLTS (deep level transient spectroscopy)}
                 \rcgindex{\myidxeffect{D}!Deep level transient spectroscopy (DLTS)}

\begin{figure}[t]
\sidecaption[t]
\includegraphics[width=7.5cm,clip]{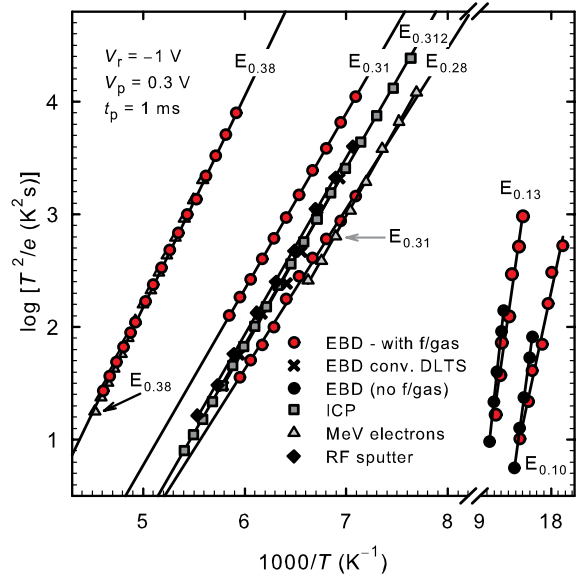}
\caption{DLTS Arrhenius plots of some electron trap defects
observed in Ge. The E-center, here marked as E$_{0.38}$ figures
among them. Reproduced with permission from: Coelho, S.M.M.,
Auret, F.D., {Janse van Rensburg}, P.J., Nel, J.: Electrical
  characterization of defects introduced in n-{Ge} during electron beam
  deposition or exposure.  J. Appl. Phys. \textbf{114}(17), 173,708 (2013).
  Copyright (2013) by AIP Publishing LLC}
 \label{germanium-figure08} 
\end{figure}

The technique known as DLTS, deep level transient spectroscopy,
uses a pn junction or a metal-semiconductor junction known as a
Schottky diode. A voltage pulse is sent through the junction in
reverse bias, so as to flood all the traps with electrons, which
after the pulse  start to emit electrons towards the conduction
band at a rate given by Eq.~(\ref{germanium-eq:emissionrate}). The
capacitance of the junction depends on the charge accumulated in
the traps and therefore changes with time as the traps become
depleted. It is measured at two different times $t_1$ and $t_2$.
If $C_0$ is the capacitance at $t_1$ and $\Delta C$ the change in
the capacitance between $t_1$ and $t_2$, it can be demonstrated
that the relative change in the capacitance $\Delta C/C_0$ has a
maximum when the so called rate window equals the emission
probability:
\begin{equation}
RW\equiv\frac{\ln(t_1/t_2)}{t_1-t_2}=e_n\,.
\label{germanium-eq:ratewindow}
\end{equation}
Typical rate windows are 80\,s$^{-1}$ and 200\,s$^{-1}$.
Measurements of the DLTS signal $\Delta C/C_0$ are performed while
the temperature $T$ is changed. When the RW equals the emission
rate of some defect a peak appears in the plot of $\Delta C/C_0$
with respect to T. In this way the different defects appear. At
the peak
\begin{equation}
N_T=2\left(\frac{\Delta C}{C_0} \right)_\mathrm{peak} N_D,
\end{equation}
where $N_D$ is the number of dopants in an n-type semiconductor
and $N_T$ is the number of traps corresponding to the peak.  Using
several RWs, several values of $e_n$ can be obtained for different
temperatures, being $E_T$ the slope of the representation
$\ln(T^2/e_n)$ with respect to $1/T$. From the same representation
the value of $\sigma_\mathrm{app}$ can be obtained and therefore
the defect is fully characterized. From the height of the peak the
concentration of the defect $N_T$ can also be obtained. The value
of the reverse bias determines the depth of the measurements and
allows for the plotting of the profile of $N_T$ as a function of
the depth of the sample. This procedure to characterize the
E-center in Ge was performed in Ref.~\cite{germanium-coelho2013}
and the Arrhenius plots for several defects can be seen in
Fig.~\ref{germanium-figure08}.

\section{Experiment of plasma-induced annealing}
\label{germanium-sec:experiment}
\begin{figure}[b]
 \begin{center}
\includegraphics[width=9cm,clip]{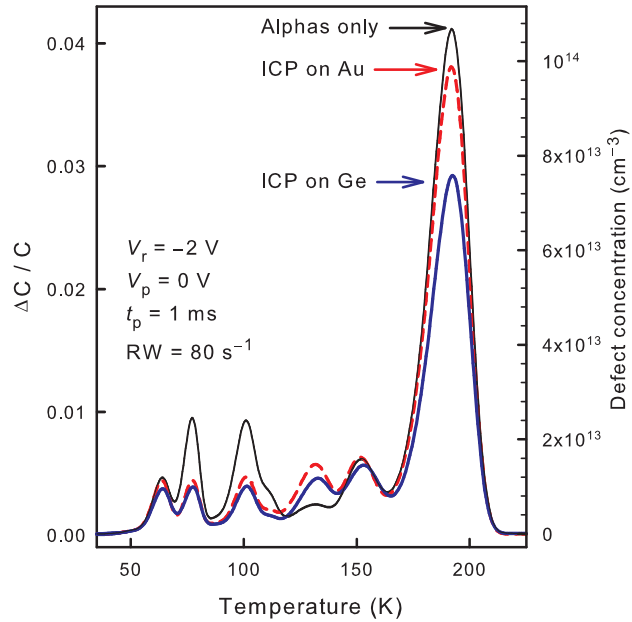}
\caption{DTLS spectra showing the experimental results. The defect
concentrations on the right axis are only valid for the peaks. The
main peaks correspond to the E-center defect. A 30\% diminution of
the concentration of this defect can be observed after 30 minutes
under the action of an inductively coupled plasma (ICP) with 4\,eV
Ar ions. If the ICP is applied through the Au contact the
diminution exists although it is substantially smaller. Reproduced
with permission from: Archilla, J.F.R., Coelho, S.M.M., Auret,
F.D., Dubinko, V.I., Hizhnyakov, V.:
  Long range annealing of defects in germanium by low energy plasma ions.
Physica D \textbf{297}, 56--61 (2015). Copyright (2015) by
Elsevier }
 \end{center}
\label{germanium-figure09} 
\end{figure}
                        \rcgindex{\myidxeffect{P}!Plasma-induced annealing of defects}
                        \rcgindex{\myidxeffect{A}!Annealing of defects by plasma}

The main experiment is done as follows: (a)~The Ge wafer is
bombarded with 5\,MeV alpha particles during 30 minutes and it is
left for 24 hours at room temperature for the defects to stabilize
as initially there is a fast
kinetic~\cite{germanium-fagepedersen2000}; (b)~The surface of Ge
is divided into two parts $A$ and $B$, then a diode is made using
resistive evaporation of Au on part $A$ and DTLS is performed to
measure the defect concentrations, (c)~The Ge sample is introduced
into an inductively coupled plasma (ICP) with 4\,eV Ar ions and
pressure of 0.1~mb for half an hour in intervals of 10 minutes to
allow for cooling; (d)~DTLS is performed in part $A$, where ICP
has been done through Au, (e) A diode is evaporated on part $B$,
where ICP has been applied directly on the Ge surface, and then,
DTLS is performed there.

The short time of alpha irradiation is done to allow for better
DLTS measurements. A concentration of about 10\% of $N_D$, as the
was obtained, or less  is ideal

The results of the three measurements are presented in
Fig.~\ref{germanium-figure09}. We will concentrate on the most
abundant defect, the E-center. (1) The concentration after alpha
damage and 24 hours rest is $N_T=1.07\times
10^{14}\,\mathrm{cm}^{-3}$; (2) After direct ICP on germanium  it
is reduced by 30\%;  (3) If the ICP is applied through the Au
contact, the reduction is about 7\%, smaller but still
significant.

Other details of interest are: (4) The sample heats up to about
40$^\circ$C in spite of the cooling intervals; (5) If there is no
cooling the sample heats up to about 65$^\circ$C and the decrease
in the rate of annealing is dramatic; (6) The defects are annealed
up to a depth of 2600\,nm inside the Ge
sample~\cite{germanium-archilla-coelho2015}; (7) If other metals
are used for the contact the annealing  also takes place as with
Au but the effect depends on the metal used; (8) If a plasma of
larger energy 8\,eV is used the annealing rate increases, but
given that a plasma of larger energy also has a larger flux, the
effect per Ar ion is much smaller (see below); (9) The temperature
to achieve a similar rate of annealing is  150$^\circ$C as deduced
in Sect.~\ref{sec:thermal} and by other
authors~\cite{germanium-markevich2004}.

There was no measurable concentration of Ar after ICP which
discards Ar channelling. Other explanations were considered and
discarded in Ref.~\cite{germanium-archilla-coelho2015} such as
multivacancy production, production of minority charge carriers,
production of defects that could diffuse and interact with the
E-center and diffusion of H that could passivate the vacancies in
the E-center.
\section{ILM hypothesis}

In this section we analyze the experiment and examine the
possibility that Ar ions produce intrinsic localized modes that
travel in a localized way with little dispersion through the
semiconductor and are able to anneal the defects. The exact nature
of these ILMs is not yet known but here it is assumed that they
have a vibrational part due to their origin from an Ar ion hit. If
they have also some charge or other properties is unknown and not
necessary for this hypothesis.

 \label{sec:hypothesis} The rate of ion-induced annealing is given by
 the following
equation:
\begin{equation}
\frac{\D N_T}{\D t}=-\sigma_i \Phi_i N_T\, ,
 \label{germanium-eq:ionrate}
\end{equation}
where $\sigma_i$ is an effective cross-section for defect
annealing  by plasma ions. It is as if imaginary Ar ions would
penetrate Ge and anneal a defect but at this stage there is no
need of an hypothesis, $\sigma_i$ is just the probability per unit
time and unit flux of Ar ions that a defect is annealed.
Integrating the equation above we obtain:
\begin{equation}
N_T(t)=N_T(0)\E^{\displaystyle -\sigma_i \Phi_i t}\,  \quad
\mathrm{or}\quad \sigma_i=-\frac{1}{\Phi_i
\,t}\ln\frac{N_T(t)}{N_T(0)}
 \label{germanium-eq:sigmai}
\end{equation}

For the experiment described with pressure $p=0.1\,$mb, that
corresponds to 4\,eV ions, the flux is
$\Phi_i=5.58\times10^{10}\,\mathrm{cm}^{-2}\mathrm{s}^{-1}$~\cite{germanium-archilla-coelho2015},
$t=30\times60$\,s and $N_T(t)/N_T(0)=0.7$, and $\sigma_i\simeq
35.6$\,\AA$^2$ is obtained. This value should be compared with
$\sigma_0=(n_\mathrm{Ge})^{-2/3}\simeq 8$\,\AA$^2$, that is, the
average area corresponding to an atom of Ge at the surface of the
semiconductor, then $\sigma_i\simeq 4.4\sigma_0$. This result
indicates that the process has an enormous efficiency. It has to
be considered with caution as also neutrals may be arriving at the
semiconductor surface, but it should not change the result by more
than one order of magnitude, probably by around  a factor of two
in the flux.
                           \rcgindex{\myidxeffect{N}!Number of ILMs}
                           \rcgindex{\myidxeffect{I}!ILMs (number of)}

                \rcgindex{\myidxeffect{A}!Annealing cross section}
                \rcgindex{\myidxeffect{C}!Cross section for annealing}
It is interesting to see what  the change in efficiency is when an
8\,eV plasma is used. The flux in this case is
$\Phi_i(8\,\mathrm{eV})=1.35\times10^{13}\,
\mathrm{cm}^{-2}\mathrm{s}^{-1}$~\cite{germanium-archilla-coelho2015}
and using only  600\,s time the concentration is reduced to 80\%
of the original. The cross section becomes
$\sigma_i(8\,\mathrm{eV})\simeq 0.26$\,\AA$^2\simeq
0.033\sigma_0$. Therefore a larger energy per Ar ion does not
increase the efficiency of the ion-annealing process but reduces
it  by  a factor of $\simeq 140$. This is coherent with our
hypothesis that the Ar$^+$ impacts produce ILMs, because ILMs have
a definite range of energies. More energy than what is required
will be dispersed into phonons which would interfere with the
propagation of the ILMs.
                   \rcgindex{\myidxeffect{A}!Annealing efficiency by ILMs}
                   \rcgindex{\myidxeffect{E}!Efficiency of annealing by ILMs}
                              \rcgindex{\myidxeffect{I}!ILM annealing efficiency}
It is also interesting to be aware of a few magnitudes to
appreciate what could be happening in the semiconductor. Suppose
that ILMs travel at a speed of the order of magnitude of the speed
of sound in Ge, $c_s=5400$\,m/s, the time needed for an ILM to
travel the measured depth $d=2600$\,nm is $\delta t=0.5$\,ns. This
means that the area for an Ar$^+$ hit in $\delta t$ is a circle
with a radius of about $10^{6}$ lattice units, or in other words
each impact and travel is completely isolated.

Note also that the traps are almost isolated as
$(N_T)^{-1/3}\simeq 2200$\,\AA\, or 370 lattice units. Therefore
there is no influence between them.

Let us introduce a couple of parameters, $\gamma$ the efficiency
of ILM creation by Ar ions, that is
\begin{equation}
\Phi_\mathrm{ILM}=\gamma\Phi_i\,
\end{equation}
and $\alpha$ the cross section for ILM defect annealing measured
in $\sigma_0$ units, that is
\begin{equation}
\sigma_\mathrm{ILM}=\alpha\sigma_0\,.
\end{equation}
Therefore
\begin{equation}
\sigma_i=\alpha\gamma \sigma_0\,
\end{equation}
and $\alpha\gamma\simeq 3.6$. The cross section should be larger
than $\sigma_0$ because the size of an E-center is at least two
atoms and due to the complex nature of Ge, ILMs probably have also
a complex structure with a few atoms involved perpendicular to the
movement of the ILMs. If the interaction takes place at a distance
of four atoms then $\alpha\simeq 8^2\sigma_0$ and $\gamma=0.06$.
The latter result implies that about 20 Ar$^+$ hits are necessary
to produce an ILM.
The number of Ar$^+$ to anneal a defect can also be  calculated
easily as $\Phi_i\,t/(0.3\,N_T\,d)\simeq 1.2\times10^4$.

In the following section it will be made clear that this rate of
annealing cannot be produced only by the increase in temperature.
Therefore, although the numbers are approximate and many
objections can be made there are a few clear consequences of this
analysis: (1)~Some entity which we call ILM, and most likely it is
a vibrational entity, is able to travel distances of a few
micrometers inside Ge in a localized way and without losing much
energy (2)~There is a high efficiency in the conversion of Ar$^+$
hits to ILMs; (3)~There is a high efficiency for ILMs to anneal or
modify defects.

Note that if the annealing barrier is $E_A$ it is neither
necessary for an ILM to have  nor to deliver $E_T$ to anneal the
defect. The change of the barrier due to the passing of an ILM
nearby brings about a change in the annealing rate which can be
very high. See
Refs.~\cite{germanium-dubinko2011,germanium-dubinko2013,germanium-coelhoarchillaquodons2015,%
germanium-dubinkoarchillaquodons2015}.

\section{Thermal annealing}
\label{sec:thermal}
 In this section we review thermal annealing
and apply it to Ge in order to compare  the temperature and energy
needed to obtain the same rate of thermal annealing as with Ar
ions.
                     \rcgindex{\myidxeffect{T}!Thermal annealing}
                     \rcgindex{\myidxeffect{P}!Phonon annealing}
                     \rcgindex{\myidxeffect{A}!Annealing (thermal)}
                     \rcgindex{\myidxeffect{A}!Annealing by phonons}

Thermal annealing of defects in semiconductors is given by a first
order kinetic
 \begin{equation}
  \frac{\D N_T}{\D t}=-K N_T\, ,
 \label{germanium-eq:thermalkinetics}
\end{equation}
where $K$, known as the reaction rate constant is given by an
Arrhenius type law
 \begin{equation}
K=A\E^{-\displaystyle E_a/k_B T}\,,
 \end{equation}
where $E_a$ is known as the annealing energy and $A$ as the
pre-exponential factor. $E_a$ can be interpreted as the potential
barrier which is necessary to surmount in order that the
transformation or diffusion process for annealing takes place. The
exponential term can be seen as the probability for an
accumulation of energy of magnitude $E_a$. The pre-exponential
term $A$ has units of frequency and it is also known as the
frequency factor. It is related to the number of attempts per unit
time that the system tries to pass the barrier and with the
curvature of the energy with respect to the reaction coordinate.
$A$ may also depend on the temperature but in a much weaker way
than the exponential term. It also depends on the entropy change.
                        \rcgindex{\myidxeffect{R}!Rate constant for annealing}
                           \rcgindex{\myidxeffect{A}!Annealing rate constant}
                    \rcgindex{\myidxeffect{A}!Activation barrier}
\begin{figure}[!t]
 \begin{center}
\includegraphics[width=6.0cm]{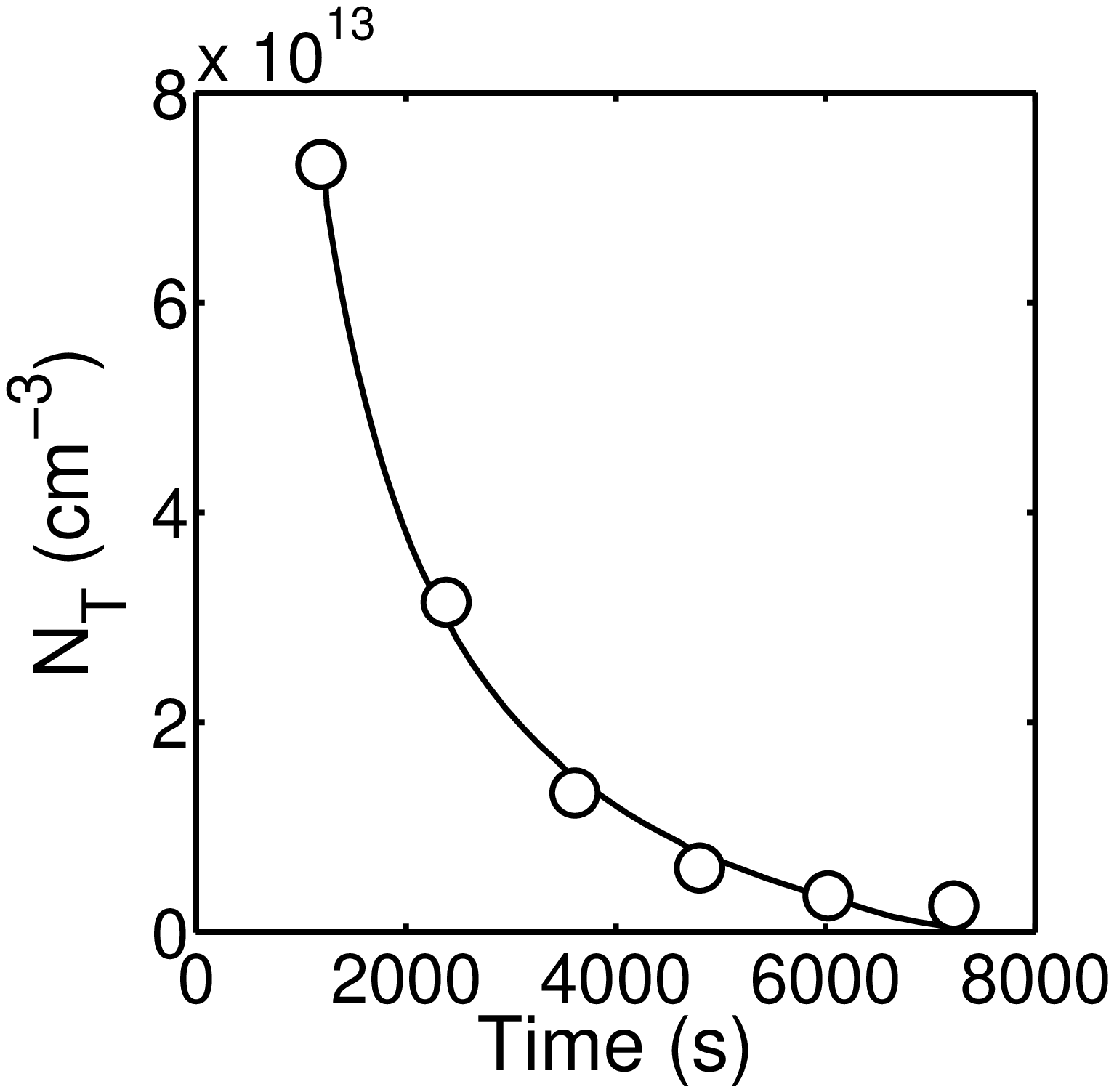}
\includegraphics[width=6.4cm]{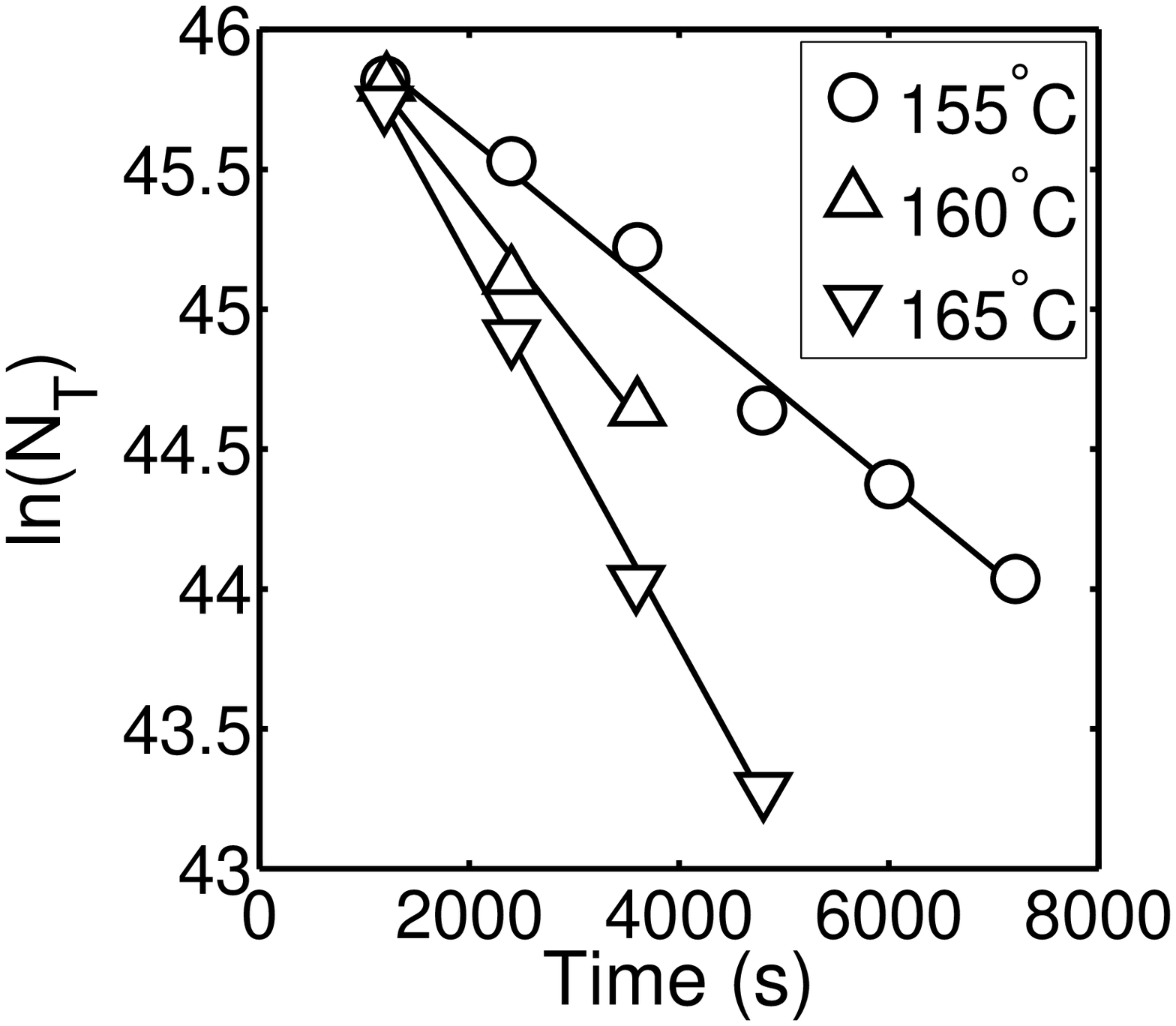}%
\\(a) \hfill \mbox{}\hfill(b) \hfill\mbox{}\\
  (c)
  \includegraphics[width=6.2cm]{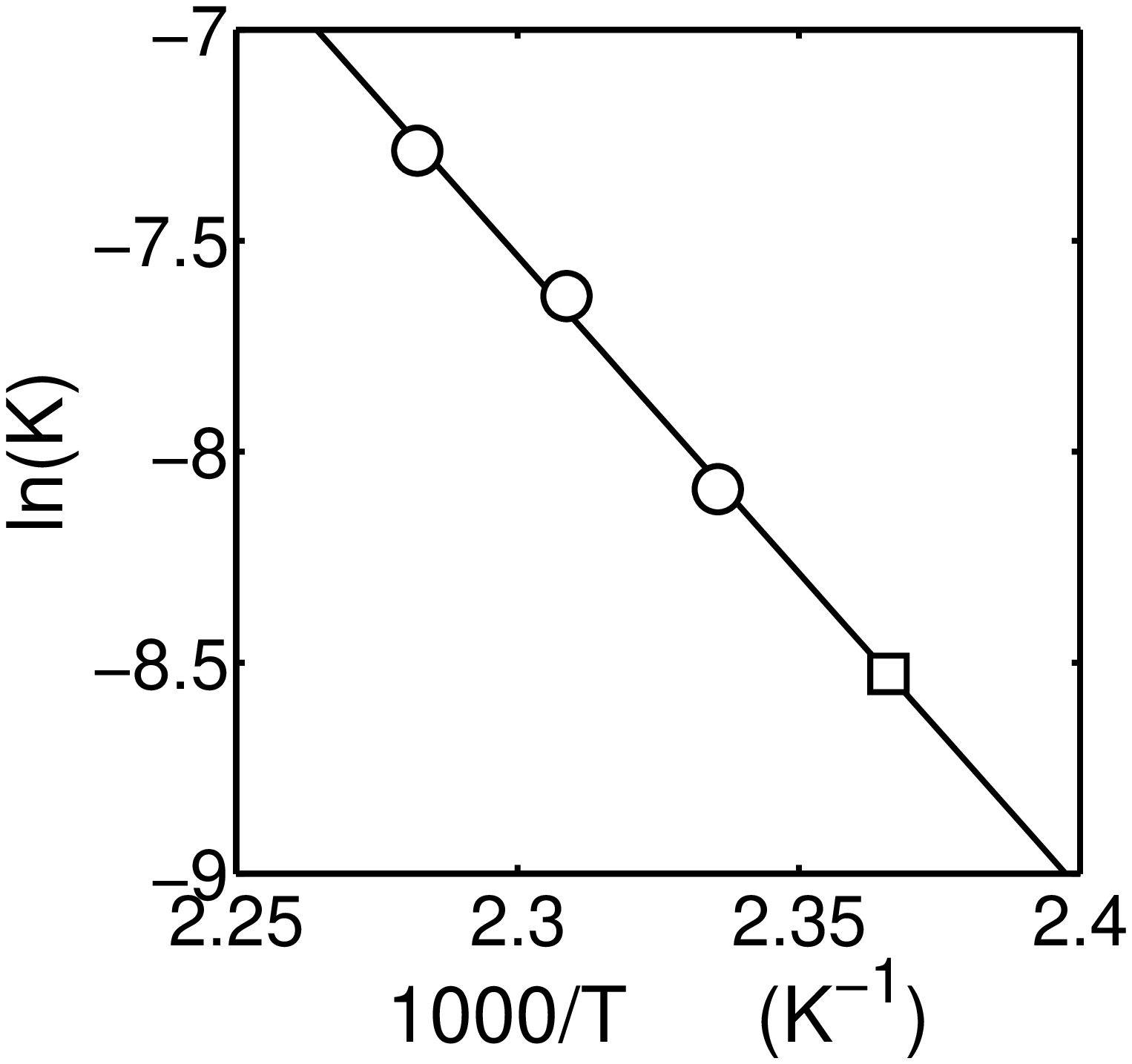}
  \caption{(a)~Defect concentration versus annealing time at $T=165^\circ$C.
  (b) Semi-log plot of defect concentration
versus annealing time at temperatures 155$^\circ$C, 160$^\circ$C
and 165$^\circ$C from which the annealing rate constant, $K$, is
calculated. (c) The Arrhenius plot from which $E_A=1.3\,eV$,
$A=0.55$\,THz and $T_A=423$\,K are obtained. Lines are fitted
curves, circles and triangles are experimental values, the square
in (c) corresponds to a thermal annealing rate equal to
ion-induced annealing. Details of the experimental procedure used
can be read in Ref.~\cite{germanium-nyamhere2009}}
 \end{center}
\label{germanium-figure10}
\end{figure}

The integration of  Eq.~(\ref{germanium-eq:thermalkinetics}) leads
to the exponential decay  $N_T(t)=N(0)\exp(-Kt)$ and comparing the
experimental data with $\ln(N(t))=\ln(N(0)) -Kt$ it is possible to
obtain $K$. Several data for E-center annealing have been
published~\cite{germanium-fagepedersen2000,germanium-markevich2004}.
Here we will use the results obtained by some of the authors
according to the procedure described in
Ref.~\cite{germanium-nyamhere2009} and using the same dopant and
defect concentration as in this work.
Figure~\ref{germanium-figure10}(a) shows the exponential decay at
165$^\circ$C and Figure~\ref{germanium-figure10}(b) represents
$\ln(N_T)$ with respect to time for three temperatures. The
approximate linear dependence can be seen. From the slopes, three
values of the reaction rate constant are obtained and in
Fig.~\ref{germanium-figure10}(c)
  $\ln(K)$ is
represented with respect to $1000/T$ and the linear dependence can
be observed.  Comparing with $\ln(K)=\ln(A) -E_a/k_BT$  the values
$A=5.5\times 10^{11}\,\mathrm{s}^{-1}$ and $E_a=1.3$\,eV are
obtained. These numbers should be treated with caution as the
experimental procedure is very sensitive to the details of the
experimental technique. The sample has to be cooled and reheated
to measure the defect concentration.

\section{Comparison of thermal and plasma-induced annealing}
Comparing the equations for thermal annealing
Eq.~(\ref{germanium-eq:thermalkinetics}) and ion-induced annealing
Eq.~(\ref{germanium-eq:ionrate}) we can observe that if both
process have the same rate of annealing
 \begin{equation}
 K=\sigma_i\Phi_i\quad   \mathrm{or}  \quad A\E^{\displaystyle
- E_a/k_bT}=\sigma_i\Phi_i \,.
 \label{germanium-eq:bothkinetics}
\end{equation}
From this equation, the value of $T_A=423$\,K is obtained.

The thermal energy at $T_A$ per unit volume using Ge density of
states $g(E)$ from Sect.~\ref{germanium-sec:phonons} is given by
 \begin{equation}
 u_\mathrm{ph}=\int_0^{Em}\langle n\rangle  E g(E)\D E\,.
 \label{germanium-eq:uphonon}
\end{equation}
Note that the use of the Einstein model with $T_E=288$\,K leads to
very similar results. The increment in energy from room
temperature $T_R=300$\,K to $T_A=423$\,K is given by
\begin{equation}
\Delta u_\mathrm{ph}= u_\mathrm{ph}(T_A)-u_\mathrm{ph}(T_R)\simeq
2.9\,\mathrm{KJ/mol}\simeq 30.1\,\mathrm{meV/atom}\,.
\end{equation}

The energy per unit volume of energy in ILMs is given by
\begin{equation}
 u_\mathrm{ILM}=\rho_\mathrm{ILM} E_\mathrm{ILM}\,,
 \label{germanium-eq:}
\end{equation}
where $\rho_\mathrm{ILM}$ is the density per unit volume of ILMs
and $E_\mathrm{ILM}$ is the mean ILM energy. Both quantities are
unknown but we can estimate both. The maximum flux of ILMs is the
flux of ions $\Phi_i$ and the maximum energy is the energy that a
4\,eV Ar ion can deliver to  a Ge atom, 3.6\,eV. Let us suppose
$E_\mathrm{ILM}\simeq 3$\,eV and $\Phi_\mathrm{ILM}\simeq\Phi_i$.
The velocity of ILMs should be of the order of magnitude of the d
velocity of sound, $v_\mathrm{ILM}\simeq c_s=5400$\,m/s. Then
$\rho_\mathrm{ILM}\simeq\Phi_\mathrm{ILM}/v_\mathrm{ILM}\simeq
10^5\,\mathrm{cm}^{-3}$ and the ILM energy per Ge atom is
\begin{equation}
 \frac{u_\mathrm{ILM}}{n_\mathrm{Ge}}=\frac{\Phi_\mathrm{ILM}\mathrm{ILM} E_\mathrm{ILM}}{v_\mathrm{ILM}\, n_\mathrm{Ge}}
\simeq 7\times 10^{-15}\,\mathrm{meV/atom}\,.
 \label{germanium-eq:uILM}
\end{equation}
This value is so small because there is only an ILM for every
$4\times10^{17}$ Ge atoms. Therefore the ratio
$u_\mathrm{ILM}/\Delta u_\mathrm{ph}\simeq 10^{-16}$, which proves
that an enormously larger amount of energy in phonons is needed in
order to produce the same annealing effect than the Ar ions
produced. Changes in the ILM energy, their speed, the number of
them created by neutrals in the plasma and other factors cannot
change their energy density by a factor of $10^{16}$.
                       \rcgindex{\myidxeffect{N}!Neutrals in plasma}


\section{Summary}
In this work we have described an experiment in which a low
energy, low flux Ar plasma anneals defects in Sb-doped Ge up to a
significant depth below the surface. The hypothesis advanced in
Ref.~\cite{germanium-archilla-coelho2015} and continued here is
that Ar ions produce some kind of travelling localized excitation
with great efficiency. We call these entities intrinsic localized
modes or ILMs because their energy and other properties indicates
that their energy is vibrational, although this is by no means
demonstrated. Some space has been dedicated to phonons in
germanium in order to have a clear picture of them and their
energies and so doing clarify that they cannot be responsible for
the annealing effect, because the ILM energy density is much
smaller than the phonon density which produces the same annealing
rate. Also we think that the study of the dispersion relation can
bring home ideas about how to construct ILMs in Ge, which will be
the confirmation of the present hypothesis but seems to be a
daunting challenge.

The numbers are approximate, many hypotheses and estimations that
have been advanced may be incorrect, however none of these
problems can change the fact of the observation of long-range
annealing in germanium produced by Ar plasma and that ILMs are the
most promising cause.

\section*{Acknowledgments} The authors were funded by  MICINN,
project FIS2008-04848; the South African National Research
Foundation and the  European Regional Development Fund, Centre of
Excellence Mesosystems: Theory and Applications. JFRA and VD
acknowledges the Physics Institute in Tartu for their hospitality.

\bibliographystyle{spmpsci}
\bibliography{germaniumarxiv2015}
\end{document}